\documentclass[twocolumn,showpacs,amsmath,preprintnumbers,pre]{revtex4}

\usepackage{graphicx}
\usepackage{amsmath}
\usepackage{amssymb}
\usepackage{natbib}

\begin{document}

\title{Long wavelength structural anomalies in jammed systems}

\author{Leonardo E. Silbert$^{1}$}
\author{Moises Silbert$^{2,3}$}

\affiliation{$^{1}$Department of Physics, Southern Illinois
  University, Carbondale, Illinois 62901, U.S.A.}

\affiliation{$^{2}$Institute of Food Research, Norwich Research Park,
  Colney, Norwich NR4 7UA, U.K.}

\affiliation{$^{3}$School of Mathematics, University of East Anglia,
  Norwich NR4 7TJ, U.K.}

\begin{abstract}
  
  The structural properties of static, jammed packings of monodisperse spheres
  in the vicinity of the jamming transition are investigated using large-scale
  computer simulations. At small wavenumber $k$, we argue that the anomalous
  behavior in the static structure factor, $S(k) \sim k$, is consequential of
  an excess of low-frequency, collective excitations seen in the vibrational
  spectrum. This anomalous feature becomes more pronounced closest to the
  jamming transition, such that $S(0) \rightarrow 0$ at the transition point.
  We introduce an appropriate dispersion relation that accounts for these
  phenomena that leads us to relate these structural features to
  characteristic length scales associated with the low-frequency vibrational
  modes of these systems. When the particles are frictional, this anomalous
  behavior is suppressed providing yet more evidence that jamming transitions
  of frictional spheres lie at lower packing fractions that that for
  frictionless spheres. These results suggest that the mechanical properties
  of jammed and glassy media may therefore be inferred from measurements of
  both the static and dynamical structure factors.

\end{abstract}


\maketitle

\section{Introduction}

The emergence of similarities between the properties of molecular glasses,
dense colloidal suspensions, foams, and granular materials, have led to the
notion of jamming \cite{liu1} - the transition between solid-like and
fluid-like phases in disordered systems - as a manner through which one can
gain a deeper understanding of the traditional liquid-glass transition and the
fascinating, complex phenomena observed in amorphous materials in general
\cite{coniglio1}. Although there has recently been some works highlighting the
differences between the onset of rigidity in jammed matter and glassiness
\cite{kurchan8,zamponi2}, the emphasis in this work is on the commonalities
between the two \cite{ohern1,ohern3,leo14,leo17}. Here, we aim to provide a
heuristic physical picture that accounts for specific, long wavelength,
structural features that emerge in the jammed state and relate these features
to their dynamical properties.

Donev et al.~\cite{torquato6} found that in the {\em hard sphere} jamming
transition, the structure factor, $S(k)$, exhibits a linear dependence on
wavenumber $k \equiv |\bf{k}|$,
\begin{equation}
  S(k) \propto k~~\textrm{ as }~~k \rightarrow 0
\label{eq1}  
\end{equation}
This behavior of $S(k)$ suggests that the total correlation function $h(r)$,
decays as $|r^{-4}|$, at large separations $r$, as deduced from the asymptotic
estimates of Fourier transforms \cite{lighthill1,stell1,silbert5}. Similarly
it suggests a long range behavior for the direct correlation function $c(r)$.
This is indeed in contrast to standard liquid state theory for liquids whose
constituents interact via a finite range potential - hence $c(r)$ is short
ranged - which predicts, $S(k) \propto k^{2}$. This anomalous low-$k$ behavior
can be interpreted as being indicative of the suppression of long wavelength
density fluctuations due to hyperuniformity \cite{torquato7}.  Here, we
propose an alternative interpretation related to large length-scale collective
dynamics.

This paper is arranged as follows. We provide a brief overview of the
Molecular Dynamics (MD) simulations used here to generate liquid and jammed
states. We then review previous results from studies of the jamming transition
pertinent to the discussion here. This is followed by a discussion on the
relevant concepts from liquid state theories that indicate that our results
and those of Ref.~\onlinecite{torquato6} for frictionless particles are indeed
rather unusual. We then present our results for the static structure factor,
$S(k)$, at small values of the wavenumber $k$, in our jammed, model glassy
system. We then put forward a conjecture that relates the asymptotic behavior
of $S(k)$ to an excess of vibrational modes relative to the Debye model. We
end with results from ongoing work on frictional systems and conclusions.

\section{Simulation Model}
The computer simulations performed here are for monodisperse, {\em soft
  spheres} of diameter $d$ and mass $m$, interacting through a finite range,
purely repulsive, one-sided, harmonic potential,
\begin{equation}
V(r) = \left\{
\begin{array}{cc}
\frac{\varepsilon}{2d^{2}}(d-r)^{2} & r<d,\\ 
  0 & r>d
\end{array}
\right\}
\label{eq2}
\end{equation}
where $r=|{\bf r}_{i} - {\bf r}_{j}|$ is the center-to-center separation
between particles $i$ and $j$ located at ${\bf r}_{i}$. The strength of the
interaction is parameterized by $\varepsilon$, which is set to unity in this
study. Most of the results presented below are for frictionless particles,
with friction coefficient $\mu = 0$. We also present preliminary results for
frictional packings using a static friction model \cite{leo7} to compare
between frictionless and frictional systems. In the frictional packings the
particle friction coefficient was varied, $0.01 \leq \mu \leq 1.0$. We
simulated systems ranging in size, $1024\leq N \leq 256000$ particles, in
cubic simulation cells of size $10d \lesssim L \lesssim 50d$, with periodic
boundary conditions, over a range of packing fractions, $\phi = N\pi
d^{3}/L^{3}$.

The jamming protocol implemented here to generate zero-temperature jammed
packings is similar to other soft-sphere protocols \cite{makse1,makse2}.
Initially, starting from a collection of spheres randomly placed in the
simulation cell at low packing fraction $\phi_{i} = 0.30$, we compressed the
system to a specified over-compressed state, $\phi = 0.74$, minimizing the
energy of the system in a steepest descent manner. At this value of $\phi =
0.74$, all the particles experience overlaps with several other particles -
their contact neighbors. The packings are mechanically stable and disordered
with no signs of long range order. To generate packings at $\phi < 0.74$, we
then incrementally reduced $\phi$ in steps of, $10^{-6} \leq \delta \phi \leq
10^{-2}$, minimizing the energy after each step. This allowed us to accurately
determine the location of the jamming transition where the system unjams at a
packing fraction $\phi_{c}$ (see below), for \emph{each independent
  realization}, down to an accuracy of $10^{-6}$ in $\phi$ for $N \leq 10000$
and an accuracy of $10^{-3}$ for $N > 10000$. For the largest system,
$N=256000$, we generated four independent configurations at each value of
$\phi$.

We also studied equilibrated liquids of frictionless spheres at a
dimensionless temperatures $T = 0.01$. This value of $T$ was chosen to compare
and contrast our results between liquid and jammed states.

Our main tool of analysis is the static structure factor $S(k)$, which we
obtain in the standard way as a direct Fourier transform of the particle
positions,
\begin{equation}
  S(k) = \frac{1}{N}\left|\sum_{i=1}^{N} \exp(\imath {\bf k} \cdot {\bf r}_{i})\right|^{2} .
\label{eq3}
\end{equation}
Here, ${\bf k}$ is the reciprocal lattice vector for the periodic simulation
cell, which is restricted to $k \geq 2\pi / L$. Although the data is noisy at
the lowest wavenumbers, our largest packings ($N=256000$) display the
anomalous low-$k$ behavior, as described below, well above the noise
level. Throughout this study, $d=m=\varepsilon=1$, and all quantities have
been appropriately non-dimensionalized.

\section{Review}
\subsection{Jamming Transition}

Frictionless, purely repulsive, soft sphere system have been shown to undergo
a zero-temperature transition between jammed and unjammed phases as a function
of density or packing fraction
\cite{durian7,makse1,makse2,ohern2,ohern3,hecke6}. In the infinite-size limit,
this transition occurs at a packing fraction coincident with the value often
quoted for random close packing, $\phi_{rcp}=0.64$ \cite{ohern3}. Above the
jamming threshold $\phi > \phi_{c}$, the jammed state is mechanically stable
to perturbations with non-zero bulk and shear moduli, whereas below the
transition $\phi < \phi_{c}$, it costs no energy to disturb the system
\cite{ohern2}. The relevant parameter here is not the absolute value of the
packing fraction, rather the \emph{distance} to the jamming transition defined
through, $\Delta\phi \equiv \phi - \phi_{\rm c}$.  Thus, as a jammed state is
brought closer to the transition point, $\Delta\phi \rightarrow 0$, it
gradually loses its mechanical rigidity and becomes increasingly soft.
Intriguingly, this loss of mechanical stability as the jamming transition is
approached can be related to diverging length scales \cite{leo14} that
characterize the extent of soft regions that determine the macroscopic
behavior of system \cite{wyart2}. Experiments \cite{dauchot3} have also
identified growing length scales in the vicinity of the jamming transition
thus promoting the idea that the jamming transition can be considered in the
context of critical phenomena. 

In this paper, we connect the long wavelength structural features observed in
$S(k)$ to correlation lengths characterizing the typical length scale of
collective, low-frequency vibrational modes in jammed, zero-temperature,
disordered packings \cite{leo14}. In Ref.~\onlinecite{leo14} it was shown that
the approach of the jamming transition in a soft sphere packing is accompanied
by a dramatic increase in the number of low-frequency vibrational modes over
the expected Debye behavior. In traditional glasses, these excess
low-frequency modes in the vibrational density of states,
$\mathcal{D}(\omega)$, are often referred to as the \emph{boson peak}
\cite{phillips1} in reference to the peak observed when plotting
$\mathcal{D}(\omega)/\omega^{2}$. For convenience, we also employ this
language here. A detailed study of $D(\omega)$ for jammed sphere-packings and
the appearance of the so-called boson peak can be found in \cite{leo14}.

\begin{figure}[t]
  \begin{center}
    \includegraphics[width=7cm]{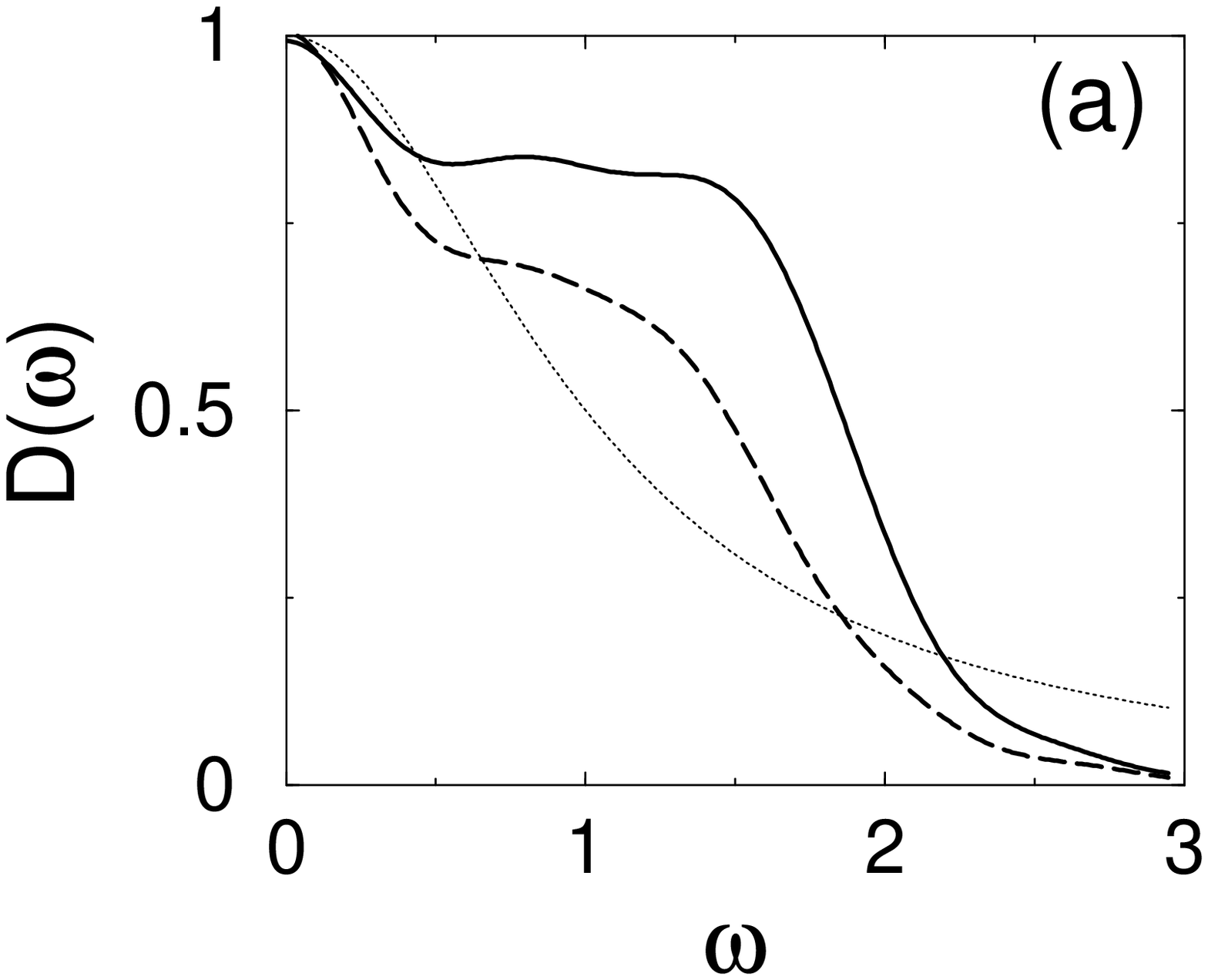}
    \includegraphics[width=7cm]{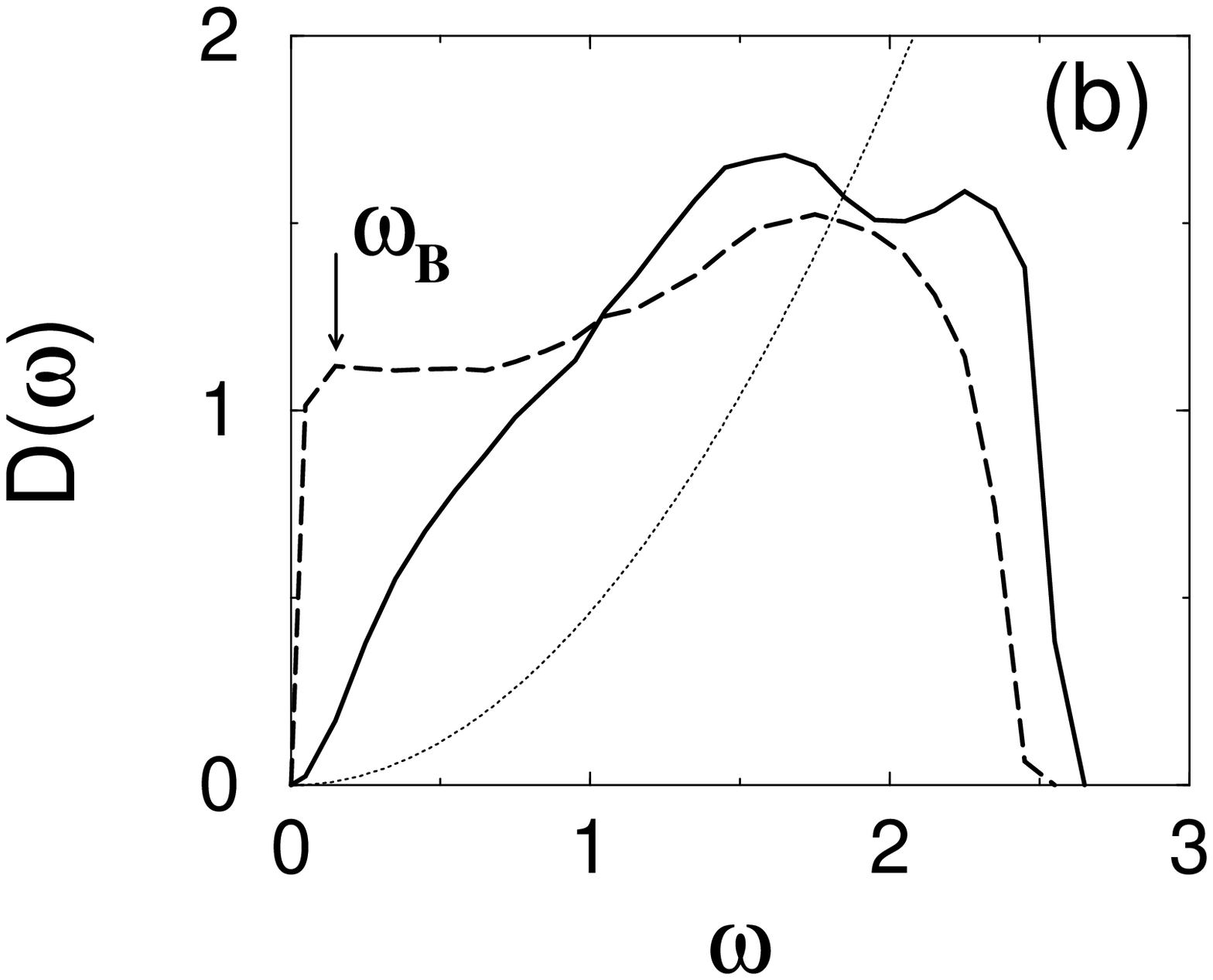}
    \caption{The vibrational density of states, ${\mathcal D}(\omega)$, for
      two systems of $N=1024$ soft, frictionless spheres, characterized by the
      distance $\Delta\phi \equiv \phi - \phi_{c}$, from their
      zero-temperature, jamming transition packing fraction. Solid line:
      $\Delta\phi=1\times 10^{-1}$.  Dashed line: $\Delta\phi=1\times
      10^{-4}$. (a) Equilibrated liquids at $T=0.01$, well above their
      respective freezing or apparent glass transition temperatures. Data
      obtained from Fourier transforms of velocity autocorrelation functions,
      and has been rescaled such that the $\omega=0$ intercepts coincide.  A
      Lorentzian function (dotted) corresponding to Langevin diffusion
      \cite{mcquarrie1} is also shown. (b) Jammed packings at zero
      temperature. The location of the \emph{boson peak} is identified as
      $\omega_{\rm B}$, for $\Delta\phi=1\times 10^{-4}$. In (b), the Debye
      result (dotted), corresponding to the system $\Delta\phi = 0.1$, was
      generated using values for the bulk and shear moduli \cite{ohern3}, and
      is shown for comparison.}
    \label{fig1}
  \end{center}
\end{figure}

In Fig.~\ref{fig1} we compare the vibrational density of states for dense,
soft sphere liquids (Fig.~\ref{fig1}(a)) and amorphous jammed solids,
identifying the location of the boson peak, $\omega_{\rm B}$, for our jammed
system (Fig.~\ref{fig1}(b)). The location of the boson peak tends to zero at
the jamming transition point, i.e. $\omega_{\rm B} \rightarrow 0$, as
$\Delta\phi \rightarrow 0$ \cite{leo14}. The two values of $\Delta\phi = 1
\times 10^{-4}$ and $1 \times 10^{-1}$ correspond to actual packing fraction
values of $\phi = 0.6405$ and $0.74$ respectively, for sample configurations
used to generate this data.

Associated with this excess in the vibrational density of states are two
diverging length scales: the longitudinal correlation length, $\xi_{L}$,
characterizing the scale of collective excitations of longitudinal modes
contributing to the boson peak, while $\xi_{T}$, characterizes transverse
excitations. These correlation lengths scale with the boson peak position as
\cite{leo14,wyart2},
\begin{eqnarray}
  \omega_{\rm B} \propto \xi_{L}^{-1}~~~{\textrm{(L: longitudinal modes)}}
  \label{eq4}\\
  \omega_{\rm B} \propto \xi_{T}^{-2}~~~{\textrm{(T: transverse modes)}}.
  \label{eq5}
\end{eqnarray}
Thus, when the Debye contribution, $\omega_{D}$, is included the corresponding
dispersion relations become,
\begin{eqnarray}
  \omega_{L}(k) - \omega_{D,L}\cong \beta k
  \label{eq6}\\
  \omega_{T}(k) - \omega_{D,T} \cong \alpha k^{2} .
  \label{eq7}
\end{eqnarray}
The preceeding relations are only approximate, valid for low frequencies. They
both contribute to the boson peak: The longitudinal term by modifying the
slope in the Debye relation, whereas the transverse contribution contains an
anharmonic term. We discuss the relevance of these results further below.

\subsection{Analyticity of $S(k)$}
In order to avoid any misunderstanding we define the concepts of regular and
singular in the following sense \cite{lighthill1}. Assuming $k$ to be a
complex variable, for any three dimensional system interacting with tempered
pairwise, additive potentials, namely when
\begin{equation}
\phi (r): \frac{1}{r^{3+\eta}}~~{\textrm as}~~r\rightarrow \infty~;~\eta>0
\label{eq8}
\end{equation}
and therefore integrable, $| \int d{\bf r} \phi (r)| < \infty$. Then, the
structure factor, in the long wavelength limit reads \cite{silbert5},
\begin{equation}
  S(k): A \rho |k|^{\eta} + F(k^{2})~~{\textrm as}~~ k \rightarrow 0
\label{eq9}
\end{equation}
where $\rho = \frac{N}{V}$ is the number density, $A$ is a constant that
depends on the thermodynamic properties of the equilibrium system, and $F$ is
an analytic function of $k$, regular in the neighborhood of $k=0$. Thus, the
second term on the right hand side of Eq.~\ref{eq9} is the regular
contribution to $S(k)$, whereas the first term - originating from the
potential of interaction - is the singular term.

For instance, when $\eta = 3$, as in the attractive part of the Lennard-Jones
potential, 
\begin{equation}
  S(k) = A\rho |k|^{3} + F(k^{2})
\label{eq10}
\end{equation}
a result originally derived by Enderby, Gaskill, and March \cite{enderby1}.
Thus, a linear behavior of the structure factor at small values of $k$
corresponds to $\eta=1$. Physically, this would correspond to a charge-dipole
interaction, which Chan {\it et al.} \cite{chan1} have shown can only be
present if \emph{long-ranged} dipole-dipole interactions are also included.
Thus, in the present work, the linear behavior of $S(k)$, at small $k$, does
not come from the potential of interaction, as the finite range harmonic
potential used here will only give the regular contribution to
$S(k)$. However, this does suggest that the anomalous low-$k$ behavior
originates in long-ranged correlations associated with the system.

\section{Results}
\subsection{Frictionless Spheres}
\begin{figure}[!]
  \begin{center}
    \includegraphics[width=7cm]{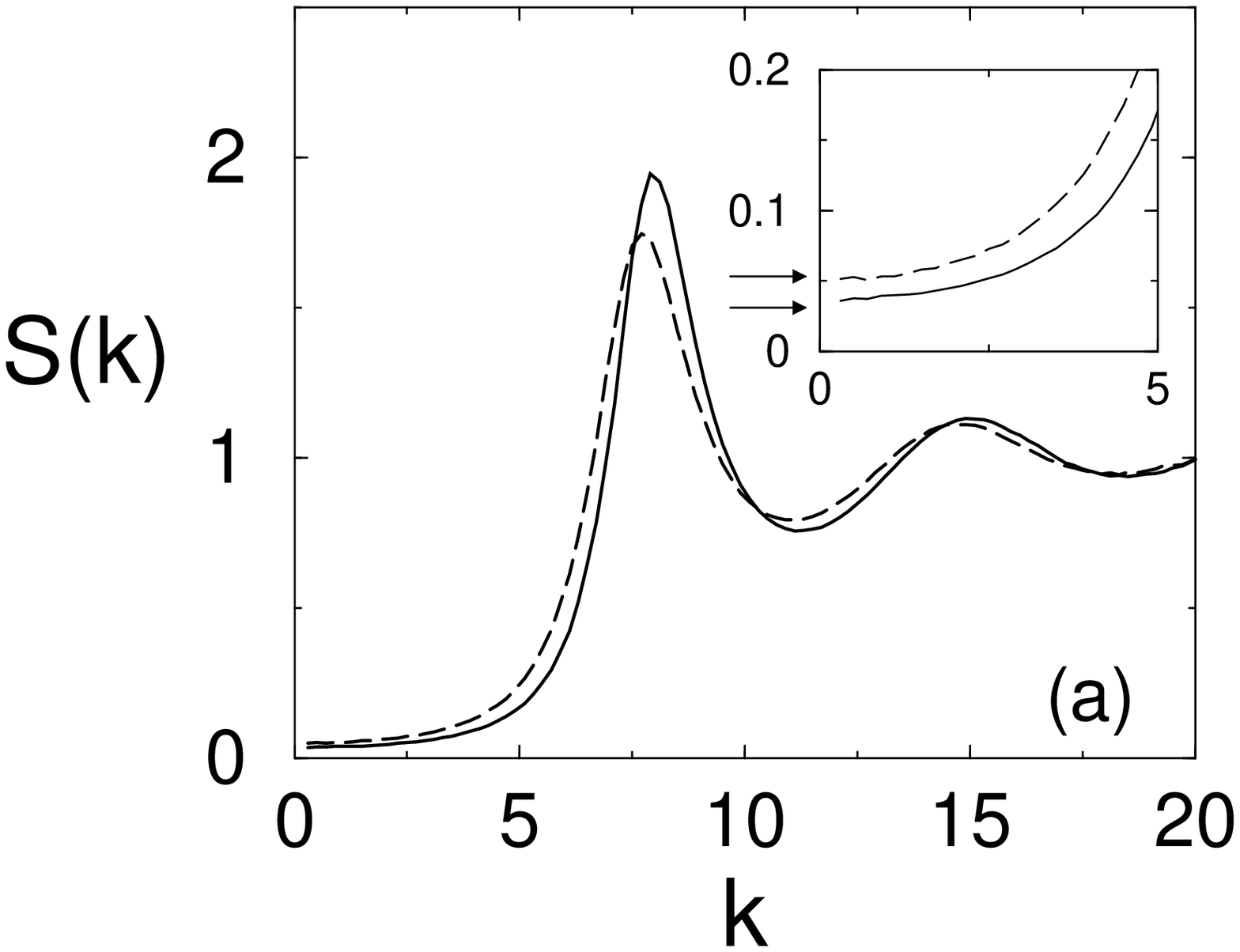}
    \includegraphics[width=7cm]{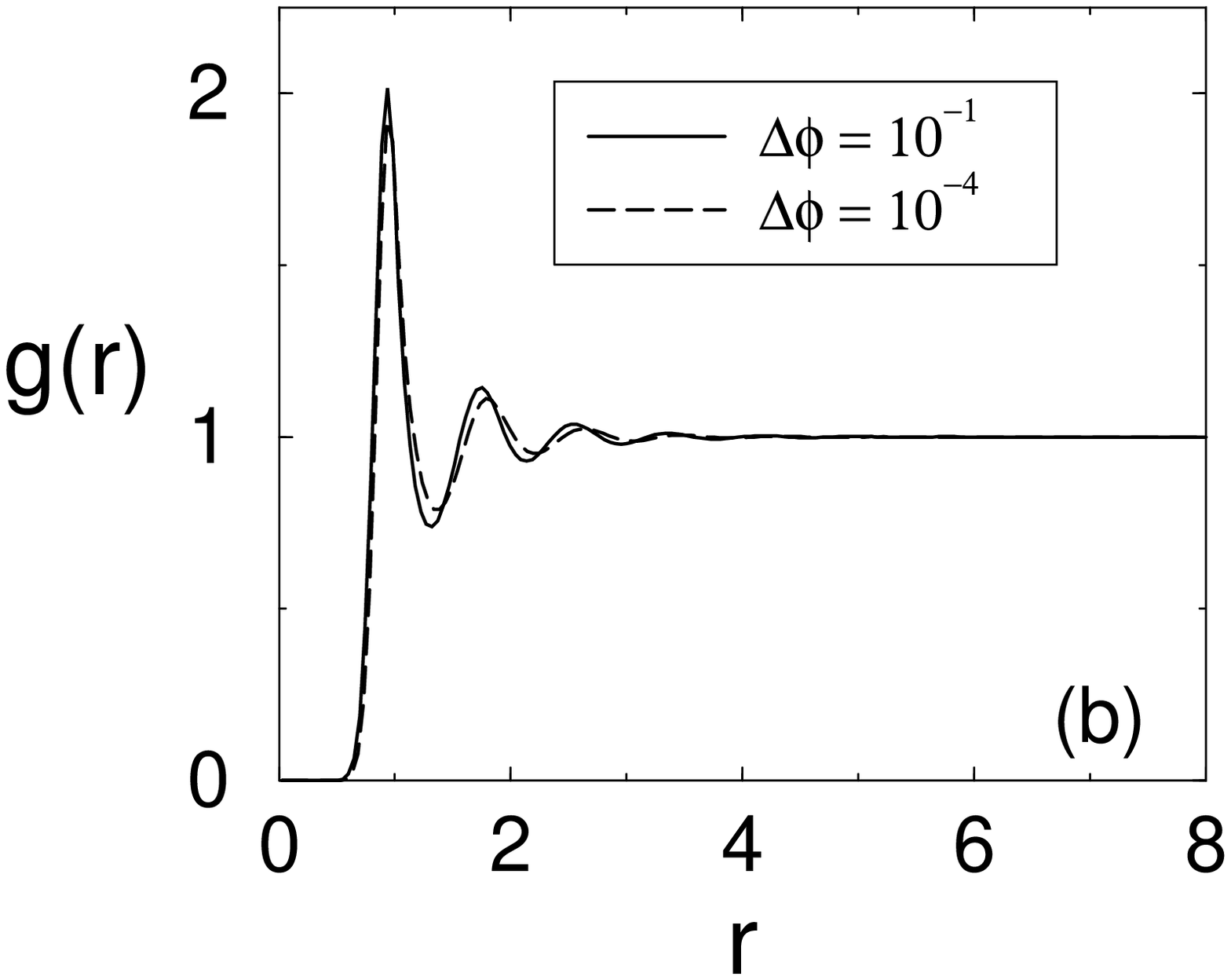}
    \caption{Two $N=10000$ soft-sphere systems in the liquid state, $T=0.01$,
      for $\Delta \phi = 1 \times 10^{-1}$ and $1 \times 10^{-4}$ (see legend
      in panel (b)). This temperature is well above the respective glass
      transition temperatures for the two systems, $T>>T_{g}$. (a) Static
      structure factor $S(k)$, as a function of wavenumber $k$. The inset is a
      zoom of the region near the origin, where the arrows indicate the
      non-zero intercepts as $k \rightarrow 0$. (b) Radial distribution
      function, $g(r)$ shows a prominent nearest neighbor peak at $r \approx
      1$ and correlations that die off rapidly.}
    \label{fig2}
  \end{center}
\end{figure}

We initially generated a number of over-compressed, $\phi > \phi_{c}$,
zero-temperature packings at various distances, $\Delta\phi \equiv \phi -
\phi_{c}$, from the jamming transition point.  We also point out that the main
distinction between this work and that of Ref.~\onlinecite{torquato6}, is that
we study soft-spheres and approach the jamming transition from above, $\phi
\rightarrow \phi_{c}^{+}$, whereas, hard spheres necessarily approach the
transition from below.  In Fig.~\ref{fig2} we show $S(k)$ for two soft-sphere
systems at $T = 0.01$, well above their respective apparent glass transition
temperatures $T_{g}$. (The corresponding density of states are shown in
Fig.~\ref{fig1}(a).) For $\Delta\phi = 10^{-1}$, $T_{g}<0.001$, while for
smaller $\Delta\phi$, $T_{g}$ lies below this. Qualitatively the curves are
similar. We find the usual primary peak corresponding to nearest neighbors,
and, as expected from standard liquid state theory \cite{mcquarrie1,silbert5},
at lower $k$, $S(k) \propto k^{2}$, with a non-zero intercept at $k=0$, which
corresponds to the finite compressibility sum rule, Fig.~\ref{fig2} inset.
Thus, for equilibrated liquids the structure factor and the long-wavelength
limit are quite insensitive to the location of the zero-temperature jamming
transition, at these values of the number density, $\rho \equiv N/L^{3} =
6\phi/\pi d^{3}= 1.22, 1.44$ for $\phi = 0.6405~(\Delta \phi = 1 \times
10^{-4}), 0.74~(\Delta \phi = 1 \times 10^{-1})$ respectively.

\begin{figure}[h]
  \begin{center}
    \includegraphics[width=7cm]{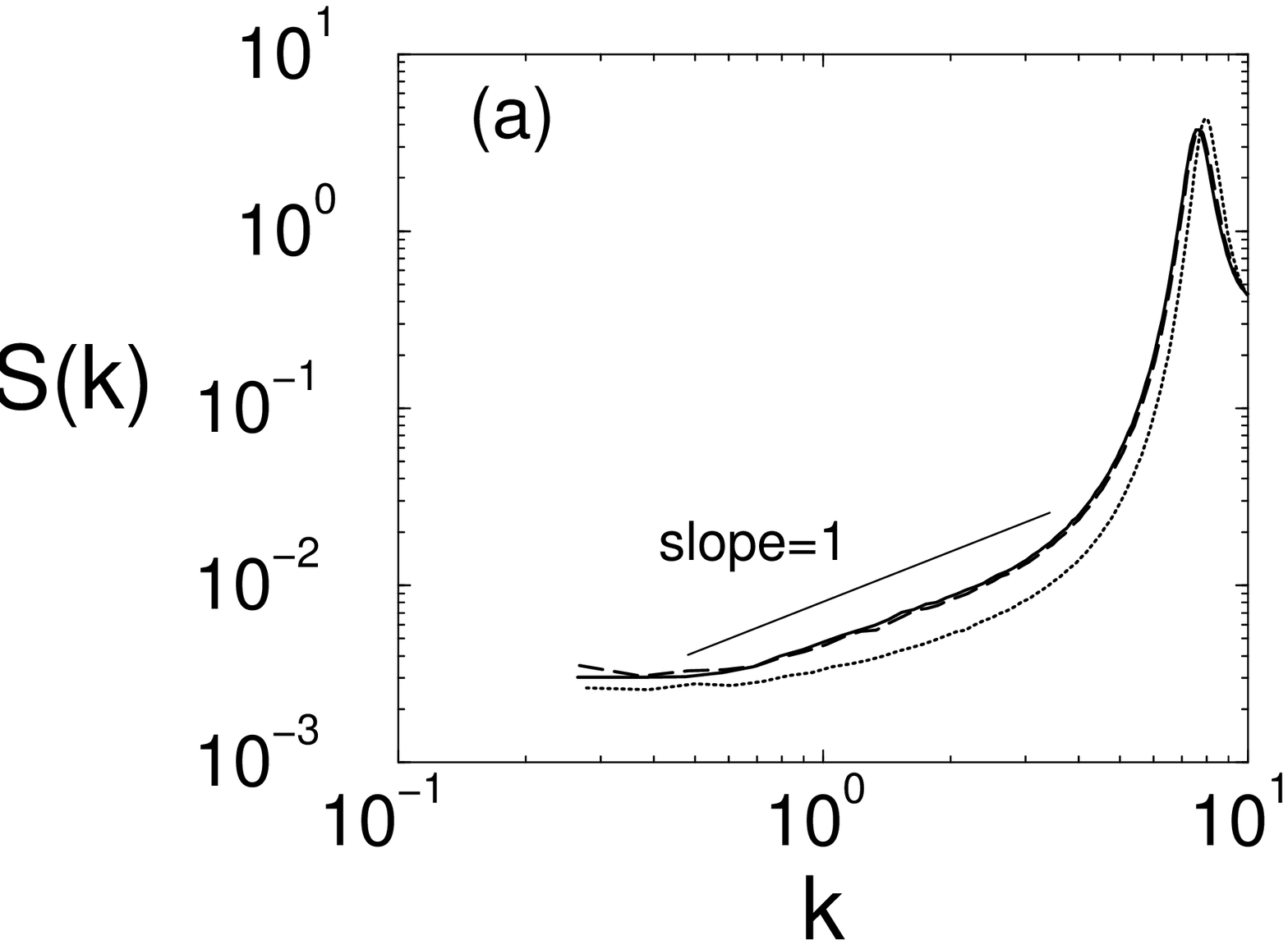}
    \includegraphics[width=7cm]{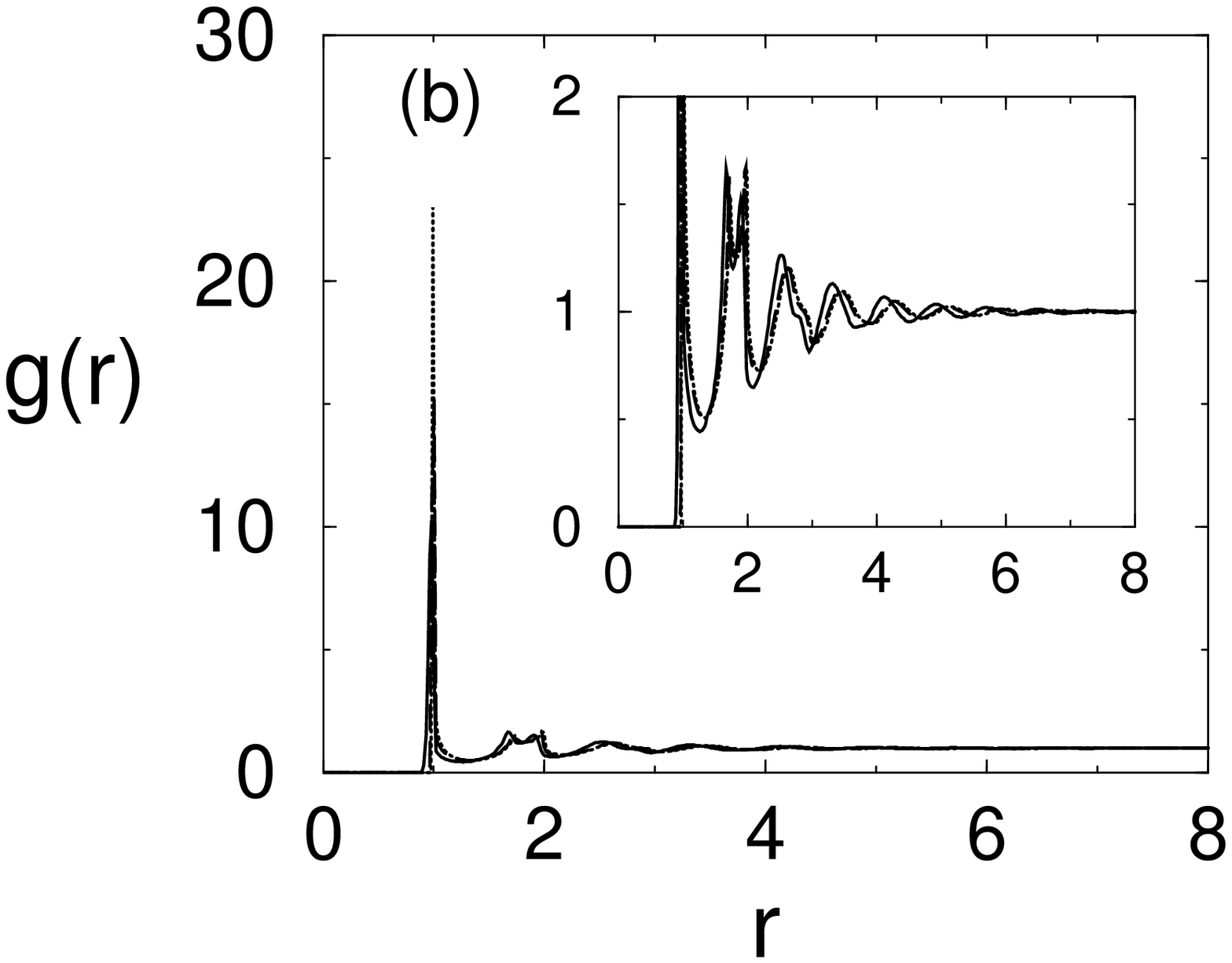}
    \caption{Zero-temperature, jammed packings containing $N=256000$
      purely-repulsive, frictionless, soft-spheres at three different values
      of $\Delta\phi = 1 \times 10^{-1}$ (dotted line), $1 \times 10^{-2}$
      (dashed), and $3 \times 10^{-3}$ (solid). (a) Static structure factor
      $S(k)$.  Far from the jamming transition, $\Delta\phi = 1 \times
      10^{-1},~S(k)$ plateaus nears $k\approx 1$.  Closer to jamming,
      $\Delta\phi = 1\times 10^{-2}$ and $\Delta\phi = 3 \times 10^{-2},~S(k)$
      exhibits approximately linear dependence on $k$ over almost an order of
      magnitude in $k$, extending down to low-$k$.  A linear curve on this
      log-log plot is shown for comparison. (b) Radial distribution
      function. The jamming transition is characterized by a diverging nearest
      neighbor peak at $r\approx 1$, a clear splitting of the second peak
      \cite{leo17}, and oscillations that persist out to larger $r$ than for
      the liquid state.}
    \label{fig3}
  \end{center}
\end{figure}

Turning our attention to the jammed phases at zero temperature
\cite{footnote23}, in Fig.~\ref{fig3} we show the radial distribution function
$g(r)$ and the low-$k$ region of $S(k)$, for $N=256000$ soft-spheres systems
at three values of $\Delta\phi = 1 \times 10^{-1}$, $1 \times 10^{-2}$, and $3
\times 10^{-3}$. Note that in Fig.~\ref{fig3}(a) we use log-log scales to
clearly demonstrate the linear region in $S(k)$. Far from the jamming
transition, $\Delta\phi = 1 \times 10^{-1}$, $S(k)$ plateaus near $k\approx 1$
and tends to constant as $k \rightarrow 0$. For the systems closer to the
jamming transition, $\Delta\phi = 1 \times 10^{-2}$ and $\Delta\phi = 3 \times
10^{-3}$, there is a \emph{qualitative change} in the low-$k$ behavior of
$S(k)$. We find a linear region, $S(k) \sim k$, extending over almost an order
of magnitude at low-$k$. We also point out that on closer inspection the
linear region extends to lower $k$ for the system closest to the jamming
transition point $(\Delta\phi = 3 \times 10^{-3})$. At the smallest $k$
attainable, $S(k)$, flattens out again because the system is not exactly at
$\Delta\phi=0$. The radial distribution function shown in Fig.~\ref{fig3}(b)
exhibits typical features characterisitc of a glassy phase; namely a split
second peak an additional shoulder on the third peak.

\begin{figure}[h]
  \begin{center}
    \includegraphics[width=7cm]{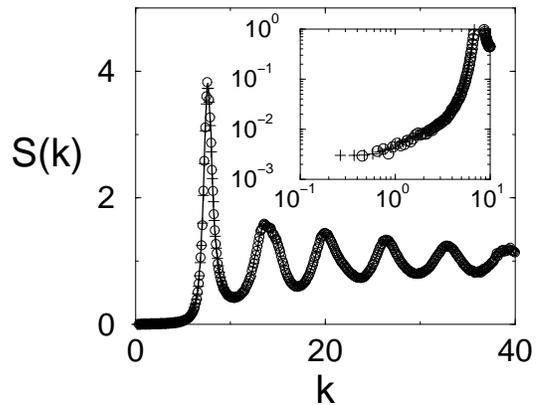}
    \caption{$S(k)$ at $\Delta\phi \approx 3 \times 10^{-3}$ for different
      systems sizes, $N = 1000$ (line), $N = 10000$ ($\circ$), and $N =
      256000$ ($+$). Oscillations persist out to large $k$, reflecting the
      dominant nearest neighbor peak in $g(r)$ shown in Fig.~\ref{fig3}. The
      inset indicates that the linear behavior at low-$k$ becomes better
      resolved with increasing system size.}
    \label{fig4}
  \end{center}
\end{figure}

For completeness, we examine the influence of system size on the results
presented here. In Fig.~\ref{fig4}, we show $S(k)$ at $\Delta\phi \approx 3
\times 10^{-3}$ for three different systems sizes, $N = 1000, ~ 10000, ~
256000$, corresponding to $L \approx 10, ~ 20, ~ 60$. The main panel of
Fig.~\ref{fig4} shows that the gross properties of $S(k)$ do not depend on
system size. Oscillations in $S(k)$ persist out to the largest $k$ and is a
consequence of the diverging nearest neighbor peak in $g(r)$ \cite{leo17}. The
inset to Fig.~\ref{fig4} indicates that the linear portion of $S(k)$ at small
$k$ becomes resolvable for $N \geq 10000$.

Thus the first aim of this work shows that, similarly to the findings in
Ref.~\onlinecite{torquato6} for hard spheres just below the jamming transition
point, for our soft sphere packings above $\phi_{\rm c}$, $S(k)$ exhibits a
linear behavior at small values of momentum transfer. Hence, this unusual
behavior not only pertains to hard spheres, but also to soft spheres, both
above and below the jamming transition point, and possibly, to finite range
repulsive potentials in general. We will attempt below to produce what we
believe is a reasonable explanation for the origin of this behavior which, as
stated at the beginning, does appeal to a conjecture.

Since, in the long wavelength limit, the linear behavior of $S(k)$
cannot arise from the singular contribution of the potential used in
this work \cite{silbert5,enderby1}, it can only be due to the
collective excitations present in the jammed or glassy state
\cite{march2}. The collective excitations we have in mind are the same
that are responsible for the boson peak. The range of $k$, over which
$S(k)\sim k$, characterizes the length scale over which these
excitations may be considered collective. As one approaches the
jamming transition, this length scale diverges, a result consistent
with work on the density of states of jammed packings \cite{leo14}.

Recent work by Chumakov {\em et al.} \cite{schirmacher2} does indeed show that
the excitations leading to the boson peak are predominantly collective; in
agreement with inelastic neutron scattering experiments \cite{ruocco1}.
Although there have been a number of theories put forward
\cite{elliott2,parisi1,wyart1}, there is, at present, no agreed explanation as
to the origins of the boson peak. For the purposes of this work this is not
necessary, except insofar as there is agreement that these excitations are
predominantly collective, and that the vibrational modes are likely to be
kinetically driven. The arguments that follow are somewhat oversimplified, but
we believe they are along the correct lines.

The following comments are in order: (i) In a system that undergoes the
liquid-glass transition, the long wavelength limit of the static structure
factor, $S(k\rightarrow 0)$, is not the thermodynamic compressibility
$\kappa_{T}$, but a different value, say $\kappa$, which can actually be
extracted from experiments \cite{gotze2}. (ii) Independently of the model or
approximation used, the density fluctuations in the glass phase are
kinetically, not thermodynamically, driven \cite{gotze2}. (iii) In spite of
their different origins, the kinetic temperature, say $T_{\textrm{dyn}}$,
defined for glasses, matches the Edwards temperature, $T_{\textrm{Edw}}$, used
in granular matter. They happen to coincide within mean field theories when
the glass is in contact with an almost zero temperature heat bath, and the
athermal grains are jammed \cite{kurchan5}.

The connection between the dispersion relations and the structure
factor, follows from the second moment of the dynamical structure
factor, $S(k,\omega)$ \cite{egelstaff1},
\begin{equation}
  2\int_{0}^{\infty} d \omega \omega^{2} S(k,\omega) = \nu^{2}_{0} k^{2},
  \label{eq11}
\end{equation}
where $\nu_{0}= \nu_{0}(T)$ (for a system in thermodynamic equilibrium
$\nu_{0}^{2} = \frac{k_{B} T}{m}$, where $k_{B}$ is Boltzmann's constant). The
static structure factor is the zero moment of $S(k,\omega)$,
\begin{equation}
  S(k) = \int d\omega S(k,\omega).
  \label{eq12}
\end{equation}

We now put forward the following conjecture: assume that there is only one
very well defined collective mode with dispersion relation $\omega_{B}(k)$,
the boson peak. This gives us a relation between the asymptotic long
wavelength behavior of $S(k)$ and the boson peak,
\begin{equation}
  S(k,\omega) = S(k)\delta(\omega - \omega_{B}(k))
  \label{eq13}
\end{equation}
whence,
\begin{equation}
  S(k) = \nu_{0}^{2} \frac{k{^2}}{2\omega^{2}_{B}(k)}.
  \label{eq14}
\end{equation}
This collective mode is associated with vibrational excitations with a
wavelength of the order of the correlation length of the jammed state. We
expand the dispersion relation at these small values of momentum transfer
\cite{march2}
\begin{equation}
  \omega_{B}(k) \cong c k + a k^{2} + ... .
  \label{eq15}
\end{equation}
In Eq.~\ref{eq15}, $c$ is the speed of sound, and the second term on the rhs
denotes departures from the usual Debye behavior, such that both $c$ and $a$
are independent of $k$. Replacing Eq.~\ref{eq15} into \ref{eq14}, in the limit
of $k \rightarrow 0$, we find,
\begin{equation}
  S(k) = \frac{\nu_{0}^{2}}{2c^{2}} \left(1 - 2\frac{a}{c}k \right) + \mathcal{O}(k^2)
  \label{eq16}
\end{equation}
Thus, on comparing Eqs.~\ref{eq15} and \ref{eq16} to Eqs.~\ref{eq6} and
\ref{eq7}, it transpires that it is the transverse modes that contribute to
the linear behavior of $S(k)$. On the other hand, the longitudinal components,
Eq.~\ref{eq6}, only contribute a constant term to $S(k)$, modifying the slope
of the Debye relation.

\begin{figure}[h]
  \begin{center}
    \includegraphics[width=7cm]{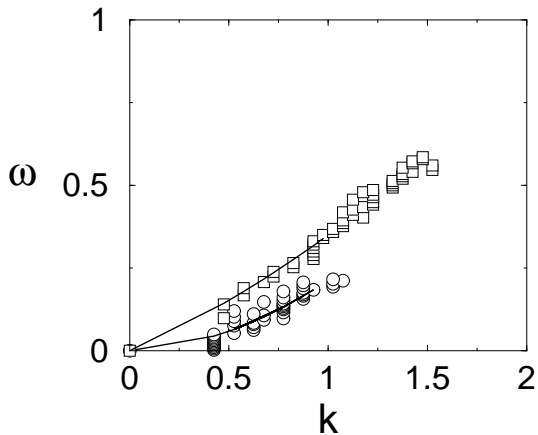}
    \caption{Transverse dispersion behavior for $\Delta\phi= 10^{-1} (\circ)$
      and $10^{-6} (\square)$. The solid lines correspond to quadratic fits to
      the data as in Eq.~\ref{eq16}. Data obtained from the low-frequency
      portion of the transverse structure factors of the vibrational modes for
      $N=10000$.}
    \label{fig5}
  \end{center}
\end{figure}

We illustrate the differences in the dispersion behavior at two extreme values
of $\Delta\phi = 10^{-6}$ and $10^{-1}$ in Fig.~\ref{fig5}. These results were
obtained from diagonalization of the dynamical matrix for systems with $N =
10000$ and then locating the peaks in the transverse components of the Fourier
transforms of the eigenmodes - transverse mode structure factors
\cite{nagel11,leo23} - and then averaging over a small range of frequencies.
The data shown in Fig.~\ref{fig5} distinguishes between the regular, linear
dispersion relation that dominates the dispersion behavior far from the
jamming transition at $\Delta\phi = 10^{-1}$, while the quadratic contribution
to the dispersion behavior is significant closer to the jamming transition,
$\Delta\phi = 10^{-6}$.

From the analysis we obtain the transverse speed of sound, $c_{t}$, as a
fitting parameter. At these two compressions: $c_{t} \approx 0.26
(\Delta\phi=10^{-1})$ and $c_{t} \approx 0.023 (\Delta\phi=10^{-6})$. The
actual values calculated from the bulk and shear moduli data \cite{ohern3}
give: $c_{t} \approx 0.28 (\Delta\phi=10^{-1})$ and $c_{t} \approx 0.018
(\Delta\phi=10^{-6})$. Therefore, we find reasonable agreement within this
approximation. If we use only a linear dispersion relation we find that the
values of $c_{t}$ computed here and the actual values \cite{ohern3} vary by as
much as an order of magnitude. This can be realized by the observation that a
linear fit to data for $\Delta \phi = 10^{-6}$ does not pass through the
origin as required in the hydrodynamic limit.

The linear feature of $S(k)$ in the small-$k$ regime appears here as a
consequence of assuming an ``excess'' relative to the Debye model for the
dispersion relation, and only one collective mode. Although the ``excess''
model used here appears as an approximation appropriate for the low-$k$
regime, we have reconciled this with existing results on the emergence of
characteristic length scales associated with these collective excitations.
More generally, however, it does suggest that the boson peak and the linear
behavior of $S(k)$, as $k \rightarrow 0$, are two sides of the same coin.
Therefore, this anomalous suppression of long-wavelength density fluctuations
emerges as a consequence of large length-scale correlated dynamics in the
low-frequency modes of the jammed solid.  Furthermore, although the boson peak
has been traditionally associated with the glassy phase, recent low frequency
Raman spectroscopic studies of glassy, supercooled, and molten silica reveal
that the boson peak persists into the liquid phase \cite{papatheodorou1}.
This also appears to be the case with other network systems (see references in
Ref.~\onlinecite{papatheodorou1}) which, in the molten state, show a
distinctive prepeak at small $k$ in $S(k)$. This prepeak is indicative of
intermediate range order in those melts, representing a characteristic length
in those melts that measures the correlations between the centers of
``clusters'' present in the liquid state. This intermediate-range clustering
may in fact promote longer wavelength correlated dynamics in the liquid state.
It may therefore be interesting to investigate experimentally, and by means of
MD simulations, whether the low $k$-behavior of $S(k)$ in those systems is
also linear, both in their glassy and liquid phases.

\subsection{Frictional Packings}

Studies on jammed packings of \emph{frictional} particles are less
well-developed. The picture that is emerging is that frictional packings
undergo a similar zero-temperature jamming transition at packing fractions
which now become friction dependent \cite{hecke7,hecke8,makse7,leo21}. In the
limit of high friction coefficient the frictional jamming transition coincides
with the value associated with random loose packing, $\phi_{rlp} \approx
0.55$ \cite{liniger1,schroter3,leo9}. Thus the relevant parameter that
measures the distance to the jamming transition now becomes a
friction-dependent quantity, $\Delta\phi(\mu)$ \cite{leo21}.
\begin{figure}[h]
  \begin{center}
    \includegraphics[width=7cm]{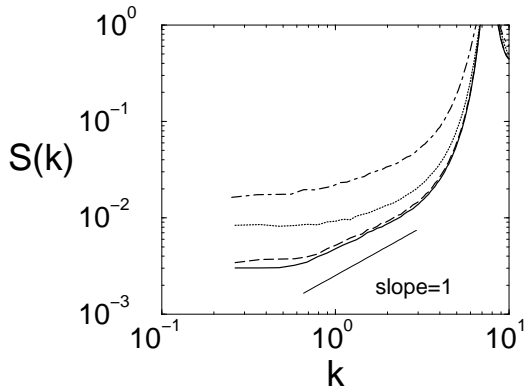}
    \caption{Log-log plot of the static structure factor $S(k)$, focusing on
      the low-$k$ region. All packings at the same $\phi = 0.64$, contain
      $N=256000$ purely-repulsive, monodisperse, soft-spheres, with different
      friction coefficients $\mu = 0~(\rm{solid-line}), ~0.01~(\rm{dash}),
      ~0.1 (\rm{dot}), ~1.0 (\rm{dot-dash})$.}
    \label{fig6}
  \end{center}
\end{figure}

We test these concepts through $S(k)$ for packings over a range of friction
coefficients at the \emph{same} $\phi = 0.64$, which translates to different
values of $\Delta\phi(\mu)$. Our preliminary results in Fig.~\ref{fig6} are
data for $S(k)$ for jammed packings with $0 \leq \mu \leq 1$. At this fixed
value of $\phi = 0.64$, $\Delta\phi(\mu)$ increases with increasing friction
coefficient. Therefore, we expect that the anomalous linear behavior in the
low-$k$ region of $S(k)$ should become less prominent with increasing $\mu$.
Indeed this trend is observed in Fig.~\ref{fig6}, thus providing yet further
evidence that the jamming transition point shifts to lower $\phi$ with
increasing friction. To recover the linear behavior at low-$k$ in frictional
packings we would therefore need to study mechanically stable packings at
lower $\phi$. This work in ongoing.

\section{Conclusions}
In conclusion, we have provided a physical, albeit naive, explanation for the
observation of the apparent suppression of long wavelength density
fluctuations in jammed model glassy materials. We submit that it is the
presence of low-frequency, collective excitations, mainly of transverse
character, contributing to the excess of low-frequency modes, that are
responsible for the linear behavior at low-$k$ in $S(k)$. The relevant length
scale here is the transversal correlation length that contributes to the
diverging boson peak in the jammed phase. This low-$k$ feature is most
pronounced in the model system studied here for soft particles at packing
fractions above the zero-temperature jamming transition matching the behavior
on the other side of the jamming transition for hard spheres. This connection
between the long-wavelength behavior on either side of the transition
reinforces the view that the jamming transition is critical in nature.
Although we also point out that the jamming transition is very different from
a thermodynamic liquid-gas critical point in that we do not observe a
divergence in $S(k=0)$, but rather it becomes zero.

Moreover, we expect that as the transition is approached,
\begin{equation}
  \nu_{0}(\mathcal{T}) \rightarrow 0 \Rightarrow S(0) \rightarrow 0 ~~~~\textrm{as}~~~~\Delta\phi \rightarrow 0.
\label{eq18}
\end{equation}
The dependence of $\nu_{0}(\mathcal{T})$ on a generalized temperature-like
quantity, $\mathcal{T}$, in jammed packings is consistent with the concept of
the \emph{angoricity} \cite{blumenfeld4,henkes3}. This plays the role of
temperature in packings of elastic particles and goes to zero when the
particles fall out of contact which occurs at the jamming transition.  Future
studies may also provide a way to understand the connection between different
temperature definitions relevant to the study of thermal and jammed systems.
Our preliminary data for frictional packings also provide further evidence
that the location of the jamming transition occurs at lower packing fractions
with increasing friction coefficient. The underlying nature of the
low-frequency modes in frictional materials has yet to be investigated in
three dimensional systems and forms part of ongoing work. When friction is
present the modes will contain not only the translational character as seen in
frictionless systems, but also rotational character due to the additional
degrees of freedom.

This anomalous low-$k$ behavior is also likely to be present in other, finite
range, model repulsive systems. These features may be detected, using the
appropriate spectroscopy, in (hard-sphere) colloidal glasses, and granular
packings, which can be prepared close to their respective jamming transitions.

\acknowledgments 

We are grateful to Gary Barker for insightful discussions on some aspects of
this work. LES is especially grateful to Jane and Gary McIntyre for support
during the course of this work and also gratefully acknowledges the support of
the National Science Foundation CBET-0828359.


\begin{thebibliography}{47}
\bibitem[{foo()}]{footnote23}
\bibinfo{note}{Because we use monodisperse spheres, we do not attempt to
  gradually cool the system towards the glass temperature, rather we analyzed
  the jammed packings that were initially generated at specified values of
  $\Delta\phi$.}
\end{thebibliography}
\end{document}